\documentclass[conference]{IEEEtran}
\IEEEoverridecommandlockouts
\usepackage{cite}
\usepackage{array}
\usepackage{multirow}
\usepackage{graphicx} 
\usepackage{tabularx} 
\usepackage{amsmath,amssymb,amsfonts}
\usepackage{algorithmic}
\usepackage{graphicx}
\usepackage{textcomp}
\usepackage{xcolor}
\def\BibTeX{{\rm B\kern-.05em{\sc i\kern-.025em b}\kern-.08em
    T\kern-.1667em\lower.7ex\hbox{E}\kern-.125emX}}
\begin{document}

\title{An End-To-End LLM Enhanced Trading System}

\author{\IEEEauthorblockN{Ziyao Zhou}
\IEEEauthorblockA{\textit{Department of Computer Science} \\
\textit{Columbia University}\\
New York, United States \\
zz2915@columbia.edu}
\and
\IEEEauthorblockN{Ronitt Mehra}
\IEEEauthorblockA{\textit{Department of Computer Science} \\
\textit{Columbia University}\\
New York, United States \\
rm4084@columbia.edu}
}

\maketitle

\begin{abstract}
This project introduces an end-to-end trading system that leverages Large Language Models (LLMs) for real-time market sentiment analysis. By synthesizing data from financial news and social media, the system integrates sentiment-driven insights with technical indicators to generate actionable trading signals. FinGPT serves as the primary model for sentiment analysis, ensuring domain-specific accuracy, while Kubernetes is used for scalable and efficient deployment.
\end{abstract}

\section{Project Overview}
Financial markets are unpredictable, and forecasting stock prices is challenging. One way to approach this is by understanding market sentiment—shaped by participants' moods and reactions. Capturing this sentiment in real time from financial news and social media is essential, but current tools often fail to provide actionable insights.

This project addresses this gap using Large Language Models (LLMs), which excel in synthesizing complex language and sentiment patterns from multiple data sources. While sentiment prediction in financial texts is a well-explored problem, traditional models have met with mixed success. By deploying an LLM-driven system, we capture nuanced sentiment in real-time, designed to support trading strategies with higher predictive accuracy.

Our end-to-end system leverages LLMs for precise sentiment extraction and tracks dynamic sentiment shifts essential for financial markets. It integrates sentiment analysis into trading strategies, allowing for practical validation through performance metrics. This comprehensive approach combines LLM-based accuracy, live multi-source data, and real-world strategy testing to offer a robust and reliable tool for informed trading decisions.

\section{Objective}
Our project introduces a novel approach to market sentiment analysis by consolidating multiple real-time data sources, including financial news and social media, into a single, comprehensive framework. Traditional sentiment tools often operate on single data sources with limited latency. By contrast, this system leverages LLMs to synthesize insights from diverse sources, offering a more holistic and timely view of market sentiment. LLMs enable the model to capture complex financial language, identifying subtle sentiment shifts that are easily missed by simpler models.

Beyond basic sentiment analysis, our system leverages LLMs for text summarization, distilling large volumes of financial news and social media data into concise, meaningful insights. This feature makes it easier for users—investors and traders alike—to quickly understand market developments without sifting through excessive information. Additionally, our approach innovatively incorporates sentiment signals and scores alongside traditional trading signals, such as SMA, RSI, and stochastic oscillators. By combining sentiment-driven insights with technical indicators, the system generates more robust and actionable trading strategies, offering a unified solution that bridges market sentiment and price-based analysis.

\section{Literature review:}
Advances in machine learning and deep learning have made sentiment analysis a focal point in NLP research, with financial sentiment analysis gaining traction through models achieving high accuracy on financial datasets. Trading-focused models, however, often rely on simpler architectures or traditional methods, lacking the nuanced understanding of LLMs and typically operating on fixed data intervals. Meanwhile, LLM-based models excel in sentiment classification but are rarely applied to direct trading contexts, where continuous, multi-source updates and trading performance validation are essential. This gap highlights the need for a real-time, adaptable solution that leverages LLM sophistication to support practical trading decisions.

Recent advances in applying LLMs for sentiment analysis have revealed that while these models capture nuanced language patterns, they can struggle to accurately differentiate sentiment intensity and direction in volatile, short-term market settings. This challenge arises because sentiment shifts in financial texts often carry subtle cues that are difficult to translate into immediate trading signals. Addressing this requires leveraging models already fine-tuned on high-frequency financial data, such as FinGPT\cite{b1}, which enhances adaptability and precision without additional training. Our approach capitalizes on these advancements by deploying a finance-specific LLM, complemented by real-time feedback loops, to align sentiment analysis with trading strategies and optimize decision-making in dynamic market conditions.

Financial sentiment analysis models have evolved significantly, beginning with FinBERT\cite{b4}, a BERT-based model fine-tuned for finance-specific language. While FinBERT improved sentiment classification in financial contexts, it was limited by batch processing, which is unsuitable for real-time trading. BloombergGPT\cite{b2} later emerged as a finance-specific transformer model, excelling in tasks like named entity recognition and sentiment analysis. However, its high costs and lack of validation in trading applications limit its accessibility and use in high-frequency trading. The latest advancement, FinGPT, demonstrates strong performance in financial sentiment tasks and is an open-source alternative; however, it remains untested in live trading environments. Our approach capitalizes on these advancements by deploying finance-specific LLMs, complemented by real-time feedback loops, to align sentiment analysis with trading strategies and optimize decision-making in dynamic market conditions.

In contrast, other event-driven stock prediction models, such as ANRES\cite{b3} and the CNN-based model\cite{b5}, incorporate sentiment data but rely on static historical time windows and limited data sources, restricting their adaptability to live market dynamics. ANRES uses an LSTM with attention to financial news, lacking real-time updates and operating on fixed dates, while Ding et al.'s model executes trades based solely on sentiment class probabilities, limiting decision-making flexibility. Similarly, the FAST model\cite{b6} employs an LSTM for sequential data processing, but it doesn’t capture nuanced contexts as effectively as LLMs and includes noisy tweet data without addressing potential inaccuracies. These limitations highlight the need for a robust, real-time model that adapts to multi-source data for continuous, dynamic trading environments.

\section{Technical Considerations}
Our primary model is FinGPT, complemented by fine-tuned versions of the Llama and Granite models for comparison on finance-specific sentiment tasks. FinGPT specializes in financial datasets and captures nuanced market language with high accuracy. Additional models help validate our findings, showing that LLMs can enhance sentiment analysis and trading strategies. This comparative approach strengthens our results and contributes to broader insights into the role of LLMs in trading.

Due to high demand on the Google Cloud Platform, we were unable to secure consistent access to GPU resources and instead utilized Colab's A100 GPUs for validation and testing. To address the memory constraints of a single A100 GPU, we selected Cohere's summarization model for its lightweight architecture and efficiency, enabling effective sentiment analysis while maintaining manageable resource requirements.

\section{Methodology and Implementations}
\begin{figure}[h!]
    \centering
    \includegraphics[width=0.4\textwidth]{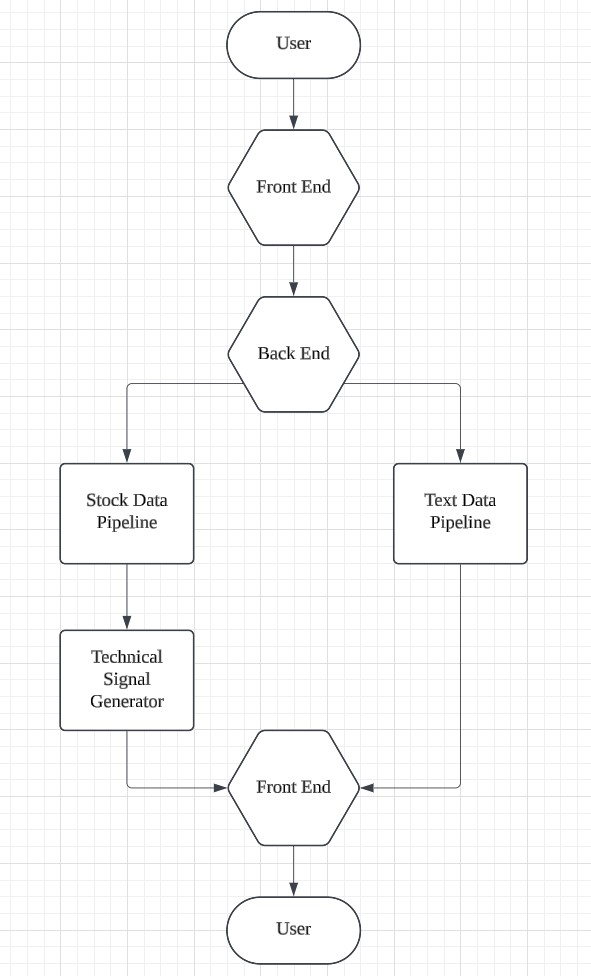}
    \caption{System Workflow Diagram}
    \label{fig:workflow}
\end{figure}

The system follows a modular workflow (Figure~\ref{fig:workflow}) to process real-time financial data, extract sentiment, and generate actionable trading signals. Users submit stock tickers through the front-end, which triggers data collection. Stock prices are aggregated into minute-level VWAP, and financial news and Reddit posts are cleaned, summarized, and analyzed using FinGPT. The processed data is combined with technical indicators—SMA crossover, RSI, and Stochastic Oscillator—to generate buy/sell signals. Results, including VWAP, trading signals, and sentiment scores, are displayed on the dashboard for decision-making. The system is deployed on Kubernetes for scalability, resilience, and efficient resource management. The detailed formulas for the trading signals and performance evaluation metrics, such as Sharpe Ratio and Win Ratio, are included in the Appendix for clarity and completeness.
\subsection*{System Workflow and Implementation}

The system workflow consists of seven key steps, from user interaction to deployment. Each step is carefully designed to ensure real-time data processing, sentiment analysis, and actionable trading signal generation.

\begin{enumerate}
    \item \textbf{User Interaction and Ticker Submission}  
    The system begins with user input, where users submit a list of stock tickers through the front-end interface. This step allows dynamic updates to the tickers being tracked.  
    \textit{Implementation:}  
    \begin{itemize}
        \item The user submits tickers via the dashboard input.  
        \item The back-end API processes the list, validates tickers and updates global configurations.  
        \item All pipelines (price, text, and signals) are restarted to reflect the new tickers.  
    \end{itemize}

    \item \textbf{Data Collection}  
    The system collects live financial data for the selected tickers from multiple sources, ensuring a comprehensive view of market activity.  
    \begin{itemize}
        \item \textit{Stock Price Data:} Real-time stock price and volume data are streamed through a WebSocket connection to Finnhub's API. Prices and volumes are continuously ingested and logged into a buffer for further analysis.  
        \item \textit{Financial Text Data:}
        \begin{itemize}
            \item \textbf{News Articles:} Relevant financial articles mentioning the selected tickers are fetched via the News API.  
            \item \textbf{Reddit Posts:} Posts and comments from the WallStreetBets subreddit are retrieved using the PRAW API. Chronological alignment ensures contextual consistency between Reddit submissions and comments.  
        \end{itemize}
    \end{itemize}

    \item \textbf{Data Processing}  
    The collected raw data undergoes preprocessing to ensure it is structured, clean, and ready for analysis.  
    \begin{itemize}
        \item \textit{Stock Price Data:} Minute-level VWAP (Volume Weighted Average Price) is calculated to smooth out short-term fluctuations and mitigate latency issues. The system retains rolling VWAP buffers for each ticker using efficient data structures like \texttt{deque}.  
        \item \textit{Financial Text Data:} Raw text data is aggregated and summarized to ensure conciseness and clarity. FinGPT, a pre-trained financial LLM, processes the summarized text to output both sentiment classification (positive/negative) and logits (confidence scores), providing granular sentiment information.  
    \end{itemize}

    \item \textbf{Sentiment Analysis and Signal Generation}  
    Processed price and sentiment data are combined to generate actionable trading signals.  
    \begin{itemize}
        \item \textit{Sentiment Analysis:} FinGPT sentiment outputs (classification and logits) are integrated as sentiment signals.  
        \item \textit{Technical Signal Generation:} Trading signals are computed using:  
        \begin{itemize}
            \item \textbf{SMA Crossover:} Fast and slow moving averages to detect short-term trend reversals.  
            \item \textbf{Relative Strength Index (RSI):} Identifies overbought or oversold conditions.  
            \item \textbf{Stochastic Oscillator:} Measures price positions relative to recent highs and lows.  
            \item \textbf{Breakout:} Sentiment scores and technical indicators are combined to produce a unified buy/sell/hold signal for each ticker.  
        \end{itemize}
    \end{itemize}

    \item \textbf{Back-End Processing and Integration}  
    The back end manages real-time orchestration between data collection, processing, and front-end display.  
    \textit{Implementation:}  
    \begin{itemize}
        \item \textbf{Concurrency:} Data pipelines (stock prices, text processing, and signal generation) run concurrently using multi-threading to ensure efficiency.  
        \item The back end dynamically serves processed data, including sentiment scores and VWAP-based signals, through FastAPI REST endpoints.  
    \end{itemize}

    \item \textbf{Front-End Visualization and User Feedback}  
    The processed insights are presented to users via a live dashboard for monitoring and interaction.  
    \textit{Implementation:}  
    \begin{itemize}
        \item \textbf{Dashboard Features:}  
        \begin{itemize}
            \item Real-time VWAP updates.  
            \item Sentiment summaries and logits generated from FinGPT.  
            \item Buy/sell/hold signals based on sentiment analysis and technical indicators.  
        \end{itemize}
        \item The dashboard refreshes dynamically every few seconds, ensuring near real-time updates.  
        \item Users can log simulated trades, which are stored for performance evaluation and further feedback.  
    \end{itemize}

    \item \textbf{Deployment}  
    \begin{itemize}
        \item The deployment process for our project involved several detailed steps to ensure a seamless setup and execution. We began by containerizing the application using Docker to create a standardized environment for the project, specifying dependencies and configurations in a Dockerfile This ensured that the application could run consistently across different systems. Next, we built the Docker image and pushed it to a container registry, such as Google Container Registry (GCR), for easy access during deployment.
        \item  Afterward, we configured a Kubernetes cluster on Google Kubernetes Engine (GKE) to handle orchestration and scalability. Using a YAML configuration file, we defined the deployment specifications, including the container image, resource limits, node selectors for GPU allocation, and environment variables. We ensured compatibility with GKE’s constraints, such as selecting an appropriate GPU type supported by the Autopilot mode. The YAML file also included configurations for service discovery and networking to expose the application.
        \item  We deployed the application by applying the Kubernetes manifests via Kubectl commands, which created the necessary pods and services. Throughout the process, we encountered challenges such as pending pods due to insufficient resources or unsupported GPU types. These were resolved by modifying the resource specifications in the YAML file and restarting the deployment. We also utilized Kubernetes commands like kubectl get pods and kubectl logs to monitor the status and debug any issues.
        \item Additionally, we implemented FastAPI as the backend framework for serving the application’s API endpoints. The API was designed to handle requests for stock sentiment analysis, fetch data from news and social media sources, and perform calculations such as VWAP and trading signals. We integrated this with external APIs like Cohere, Finnhub, and NewsAPI for real-time data processing. The deployment also required setting up logging mechanisms to monitor pipeline activities and debug errors.
        \item  Finally, a dashboard was deployed for real-time data visualization, allowing users to interact with the application via HTTP endpoints. The dashboard periodically fetched data and updated metrics dynamically, providing insights such as stock sentiment, trading signals, and trading logs. This comprehensive deployment process ensured a fully functional and scalable system ready for production.
    \end{itemize}
\end{enumerate}

\section{Results and Evaluations}
\subsection{Benchmark against other models in sentiment analysis}
We benchmarked FinGPT, IBM Granite 3.0, and Meta LLaMA 3.1 using precision, recall, F1-score, and accuracy. These metrics evaluate performance by considering both correct predictions and the balance of false positives and negatives.

\begin{table}[h!]
\centering
\footnotesize 
\caption{Performance Comparison of Models}
\label{tab:performance_comparison}
\renewcommand{\arraystretch}{1.5} 
\setlength{\tabcolsep}{8pt} 

\begin{tabular}{|l|c|c|c|c|}
\hline
\textbf{Model} & \textbf{Accuracy} & \textbf{Precision} & \textbf{Recall} & \textbf{F1-Score} \\ \hline
FinGPT          & 0.7462           & 0.7675            & 0.7642          & 0.7488            \\ \hline
IBM Granite 3.0 & 0.5861           & 0.6942            & 0.5861          & 0.6207            \\ \hline
Meta LLAMA 3.1  & 0.6565           & 0.6657            & 0.6565          & 0.6440            \\ \hline
\end{tabular}
\end{table}

The dataset chosen for this benchmarking exercise, sourced from Kaggle’s Financial Sentiment Analysis dataset, contains financial news headlines and their corresponding sentiment labels. This dataset is ideal for evaluating models in a trading-related context, as it captures the nuances of financial language and the sentiment that drives market decisions. The presence of domain-specific terminology and diverse sentiment expressions makes it challenging and well-suited for assessing the robustness and adaptability of language models in extracting financial sentiment.

Among the models evaluated, FinGPT outperformed IBM Granite 3.0 and Meta LLaMA 3.1 across all metrics. The superior performance of FinGPT can be attributed to its architecture, which is specifically designed for financial applications. FinGPT employs a lightweight Low-Rank Adaptation (LoRA) fine-tuning approach, enabling it to effectively specialize on financial datasets with minimal computational overhead. LoRA fine-tuning optimizes only a subset of model parameters, introducing low-rank matrices to capture task-specific features while retaining the core pre-trained model's general knowledge. This approach not only ensures computational efficiency but also allows the model to better adapt to the domain-specific vocabulary and sentiment patterns present in the financial dataset.

IBM Granite 3.0 and Meta LLaMA 3.1, while robust in general NLP tasks, struggled in financial sentiment analysis due to their lack of domain-specific fine-tuning and generalized training objectives. The lack of targeted fine-tuning for financial data hindered their ability to capture subtle sentiment shifts in complex financial text. In contrast, FinGPT excelled by leveraging Low-Rank Adaptation (LoRA) fine-tuning, which efficiently adapted the model to capture sentiment nuances in financial jargon. This specialization allowed FinGPT to outperform its competitors in extracting precise and actionable insights, highlighting its potential for real-time decision-making in sentiment-driven trading strategies.

\subsection{Backtest Validation}
To evaluate the performance of the trading system, we conducted backtests through the historical prices and Reddits in 2022 and 2023 using two frameworks: one incorporating sentiment signals and the other using only traditional technical indicators. Both strategies follow a position-based approach, where trades are executed dynamically based on buy/sell signals. In the sentiment-enhanced strategy, trading decisions are influenced by sentiment signals, where the signal strength determines the trade size ($10\% or 15\%$ of initial cash), and trades are classified based on sentiment polarity (positive/negative). Conversely, the base strategy executes trades purely on technical indicators without considering sentiment. In both cases, position flipping (from long to short or vice versa) triggers profit-taking, and positions are closed at the last available price. Key metrics, Sharpe ratio, and win ratio are calculated to compare the effectiveness of the sentiment-integrated strategy against the baseline. This dual backtesting approach allows us to analyze the added value of incorporating sentiment signals into trading decisions. 
\begin{table}[h!]
\centering
\footnotesize 
\caption{SHARPE RATIO COMPARISON}
\label{tab:sharpe_ratio}
\renewcommand{\arraystretch}{2} 
\setlength{\tabcolsep}{8pt} 

\begin{tabular}{|l|l|c|c|}
\hline
\textbf{Ticker} & \textbf{Strategy} & \textbf{Base} & \textbf{Sentiment} \\ \hline
\multirow{3}{*}{TSLA} 
    & SMA Crossover   & 0.34  & 3.47 \\ \cline{2-4}
    & RSI             & 0.15  & 2.37 \\ \cline{2-4}
    & Stoch. Osc.     & -1.58 & 1.79 \\ \hline
\multirow{3}{*}{AAPL} 
    & SMA Crossover   & -4.03 & 2.13 \\ \cline{2-4}
    & RSI             & -1.02 & 1.58 \\ \cline{2-4}
    & Stoch. Osc.     & -2.75 & 1.61 \\ \hline
\multirow{3}{*}{AMZN} 
    & SMA Crossover   & -2.75 & 3.14 \\ \cline{2-4}
    & RSI             & -0.95 & 2.32 \\ \cline{2-4}
    & Stoch. Osc.     & -1.03 & 2.12 \\ \hline
\end{tabular}
\end{table}

\begin{table}[h!]
\centering
\footnotesize
\caption{WIN RATIO COMPARISON}
\label{tab:win_ratio}
\renewcommand{\arraystretch}{2}
\setlength{\tabcolsep}{8pt}

\begin{tabular}{|l|l|c|c|}
\hline
\textbf{Ticker} & \textbf{Strategy} & \textbf{Base} & \textbf{Sentiment} \\ \hline
\multirow{3}{*}{TSLA} 
    & SMA Crossover   & 32.2\% & 57.0\% \\ \cline{2-4}
    & RSI             & 52.3\% & 51.4\% \\ \cline{2-4}
    & Stoch. Osc.     & 70.7\% & 64.3\% \\ \hline
\multirow{3}{*}{AAPL} 
    & SMA Crossover   & 29.9\% & 54.9\% \\ \cline{2-4}
    & RSI             & 49.5\% & 50.9\% \\ \cline{2-4}
    & Stoch. Osc.     & 78.0\% & 72.1\% \\ \hline
\multirow{3}{*}{AMZN} 
    & SMA Crossover   & 30.1\% & 64.3\% \\ \cline{2-4}
    & RSI             & 49.2\% & 52.0\% \\ \cline{2-4}
    & Stoch. Osc.     & 78.5\% & 65.1\% \\ \hline
\end{tabular}
\end{table}

\subsubsection{Sharpe Ratio}
\begin{itemize}
    \item The sentiment-integrated strategies consistently outperform the baseline across all tickers and strategies.

    \item For example, the SMA Crossover strategy for TSLA improves from 0.34 to 3.47, demonstrating a notable enhancement in risk-adjusted returns.

    \item Similarly, AAPL and AMZN show significant improvements, with Sharpe Ratios turning positive, indicating reduced volatility and improved profitability.

\end{itemize}
\subsubsection{Win Ratio}
\begin{itemize}
    \item Sentiment integration also improves the Win Ratios, though the magnitude varies across tickers and strategies.

    \item For TSLA, the SMA Crossover win ratio jumps from $32.2\%$ to $57.0\%$, while similar improvements are seen in AAPL and AMZN, especially for SMA strategies.

    \item Some indicators, such as RSI and Stochastic Oscillator, show smaller or slightly mixed improvements, suggesting sensitivity to market conditions and signal reliability.

\end{itemize}
\subsubsection{Implications}
\begin{itemize}
    \item \textbf{Enhanced Predictive Power:}  
    Integrating sentiment signals helps capture market dynamics and investor sentiment that traditional technical indicators alone may miss. This improves the ability to identify profitable opportunities, especially in volatile stocks such as TSLA.

    \item \textbf{Versatility Across Strategies:}  
    While SMA Crossover shows the most substantial improvements, all strategies benefit from sentiment integration. This highlights the versatility of sentiment signals to complement various trading approaches.

    \item \textbf{Stock-Specific Sensitivity:}  
    The results reveal that sentiment signals are particularly impactful for certain tickers, such as TSLA and AMZN, which are more popular and widely discussed. Strategies for tickers with strong market reactions to sentiment will gain the most from this integration.

    \item \textbf{Practical Use for Investors:}  
    Investors can leverage sentiment-integrated signals to improve decision-making, especially when traditional indicators generate weak or ambiguous signals. This approach reduces the volatility of the strategy and increases the robustness of the trading outcomes.
\end{itemize}
\section{Potential Challenges and Future Directions}
The main challenges encountered revolve around resources and latency in real-time data processing. Data fetching, running sentiment analysis through FinGPT, and generating actionable signals are computationally expensive tasks. Operating with limited GPU resources increases latency, which can hinder real-time performance, a critical requirement for trading systems. As the system scales to handle more assets and users, ensuring efficient data flow and faster inference will remain a key focus.
Looking ahead, the system will evolve to address these challenges while introducing enhanced capabilities.
\begin{itemize}
    \item Customizable Strategies: To offer greater flexibility, a codable interface will be developed, enabling users to define and implement their trading strategies. Instead of relying solely on existing strategies, users will have the freedom to integrate custom logic, tailoring the platform to their unique trading preferences.

    \item Interactive Trading Platform: The platform will expand beyond stocks to incorporate other trending assets, such as cryptocurrencies, catering to a broader range of market participants. Furthermore, integrating broker endpoints will allow users to place trades directly through the system. Additional features, such as real-time risk monitoring, position tracking, and interactive visualizations (e.g. price charts, and candlestick charts), will transform the platform into a comprehensive and dynamic trading hub.

\end{itemize}

\clearpage
\appendix
\section{Trading Strategies and Formulas}
This Appendix provides detailed formulas for the signals and performance metrics. Our GitHub link is:https://github.com/Ronitt272/LLM-Enhanced-Trading
\subsection{Moving Average Crossover}
\begin{itemize}
    \item \textbf{Fast Window:} 5  
    \item \textbf{Slow Window:} 30  
    \item \textbf{Long Signal:} Fast EMA crosses slow EMA from below.  
    \item \textbf{Short Signal:} Slow EMA crosses fast EMA from above.  
\end{itemize}
The Moving Average Crossover strategy identifies trends using short-term (Fast EMA) and long-term (Slow EMA) exponential moving averages.  
\begin{quote}
A \textit{long position} is triggered when the fast EMA crosses above the slow EMA, signaling upward momentum. A \textit{short position} is triggered when the slow EMA crosses above the fast EMA, signaling downward momentum.
\end{quote}

---

\subsection{Relative Strength Index (RSI)}
The \textbf{Relative Strength Index (RSI)} is a momentum oscillator that measures the speed and magnitude of price changes.  
\[
\text{RSI} = 100 - \frac{100}{1 + RS}, \quad RS = \frac{\text{Average Gain (n periods)}}{\text{Average Loss (n periods)}}
\]
\begin{itemize}
    \item \textbf{Oversold (30):} Buy signal when RSI crosses \textbf{above 30}.  
    \item \textbf{Overbought (70):} Sell signal when RSI crosses \textbf{below 70}.  
\end{itemize}
The 15-minute RSI balances responsiveness with noise reduction, helping to identify potential trend reversals.

---

\subsection{Stochastic Oscillator Strategy}
The \textbf{Stochastic Oscillator} compares a security’s closing price to its price range over a given lookback period.  

\noindent \textbf{Calculation:}
\begin{itemize}
    \item \%K Line:  
    \[
    \%K = \frac{\text{(Current Close - Lowest Low)}}{\text{(Highest High - Lowest Low)}} \times 100
    \]
    \item \%D Line: A 3-period Simple Moving Average (SMA) of \%K.
\end{itemize}

\noindent \textbf{Trading Rules:}
\begin{itemize}
    \item \textbf{Long Signal:} \%K crosses \%D below the oversold level (20), signaling upward momentum.  
    \item \textbf{Short Signal:} \%K crosses \%D above the overbought level (80), signaling downward momentum.  
\end{itemize}

---

\section{Performance Metrics}

\subsection{Sharpe Ratio}
The \textbf{Sharpe Ratio} measures the risk-adjusted return of a strategy and is defined as:
\[
\text{Sharpe Ratio} = \frac{E[R_p - R_f]}{\sigma_p}
\]
Where:  
\begin{itemize}
    \item \( R_p \): Portfolio return.  
    \item \( R_f \): Risk-free rate.  
    \item \( \sigma_p \): Standard deviation of portfolio returns.  
\end{itemize}

---

\subsection{Win Ratio}
The \textbf{Win Ratio} measures the proportion of profitable trades relative to the total trades:
\[
\text{Win Ratio} = \frac{\text{Number of Winning Trades}}{\text{Total Number of Trades}}
\]

---
\end{document}